\documentclass[aip,cha,preprint,nofootinbib,longbibliography]{revtex4-1} 

\usepackage{amsmath}

\usepackage{array}
\usepackage{amssymb}
\usepackage{ulem}
\usepackage[margin=1in]{geometry}
\usepackage{graphicx}
\usepackage{subfigure}                                                                                          
\usepackage[bookmarks=false,hyperfootnotes=false]{hyperref}
\usepackage{epstopdf}
\usepackage{multirow}
\usepackage{epstopdf}

\usepackage{colortbl}
\usepackage{hhline}

\usepackage[raggedright]{titlesec}

\usepackage{tikz}
\usetikzlibrary{fit,shapes.misc}

\newcommand{\p}{\mathbb{P}}
\newcommand{\pa}{p_i^{\uparrow}}

\newcommand{\pd}{p_i^{\downarrow}}

\begin{document}

\thispagestyle{empty}

\title{Stabilizing Gene Regulatory Networks Through Feedforward Loops}
			
\email[send to]{claus89@vt.edu}

\noindent {\textbf{\Large{Stabilizing Gene Regulatory Networks Through Feedforward Loops}}}\\

\setlength{\leftskip}{1in}

\noindent{\large{C. Kadelka,$^{1,2,}$\footnote[4]{Author to whom correspondence should be addressed. Electronic mail: claus89@vt.edu} D. Murrugarra,$^3$ and R. Laubenbacher$^{1,2}$}}\\

\noindent{\small $^{1}$ Bioinformatics Institute, Virginia Tech, Blacksburg, Virginia, 24061, USA }

\noindent{\small $^{2}$ Department of Mathematics, Virginia Tech, Blacksburg, Virginia, 24061, USA }

\noindent{\small $^{3}$ School of Mathematics, Georgia Tech, Atlanta, Georgia, 30332, USA }\\


\noindent \textit{The following article has been accepted by Chaos. After it is published, it will be found at http://chaos.aip.org}\\

\noindent The global dynamics of gene regulatory networks are known to show robustness to perturbations in the form of intrinsic and extrinsic noise, as well as mutations of 
individual genes. One molecular mechanism underlying this robustness has been identified as the action of so-called microRNAs that operate via feedforward loops. We present results of a computational study, using the modeling framework of stochastic Boolean networks, 
which explores the role that such network motifs play in stabilizing global dynamics. The paper introduces a new measure for 
the stability of stochastic networks. The results show that certain types of feedforward loops do indeed buffer the network against
stochastic effects.

\setlength{\leftskip}{0pt}

\section*{}
\noindent\textbf{The term {\textit{canalization}} was coined by the geneticist C. Waddington 
to describe the theory that embryonal development is buffered against genetic and environmental perturbations.
It is only recently that a molecular basis for
this phenomenon has been suggested.
Recent research has highlighted how the intrinsic stochasticity of gene expression
can drive changes in phenotypes.
Short segments of single-stranded RNA, so-called microRNAs (miRNA), represent
an entirely novel agent of gene regulation
discovered relatively recently, and
have been proposed to function as canalizing agents that buffer the effects of such stochasticity in gene expression.
According to this theory,
when miRNA expression is perturbed, stochasticity in gene expression can result in transitions to
distinct cellular phenotypes. As miRNAs bind to gene targets they
downregulate translation of target mRNA into protein. Embedded in several different
types of so-called feedforward loops (FFLs),
miRNAs help smooth out noise and generate canalizing effects in gene regulation by overriding
the effect of certain genes on others.}

\noindent \textbf{Much experimental work remains to be done in elucidating this concept, and recent years have seen
an explosive growth of publications in this area. There have also been a number of computational studies
focused on canalization.
In this paper, we carry out a computational study of the ability of the feedforward loop motif to buffer a gene regulatory network
against intrinsic noise. This is done using stochastic Boolean network models as a computational instantiation of
gene regulatory networks. We introduce a measure on networks that
captures its ``distance-to-deterministic" characteristics in terms of the stability of their attractors. For a given network, we
successively introduce feedforward loops and track the resulting change in dynamics. The results show clearly that the
feedforward loop motif buffers the network phenotype, in terms of stability of attractors, against perturbations from
intrinsic noise. }

\section{Introduction}
The term {\textit{canalizaton}} was coined by the geneticist C. Waddington \cite{wad}
to describe the theory that embryonal development is buffered against genetic and environmental perturbations. 
It is only recently that a molecular basis for
this phenomenon has been suggested. 
Recent research has highlighted how the intrinsic stochasticity of gene expression 
can drive changes in phenotypes \cite{Elowitz}. 
Short segments of single-stranded RNA, so-called microRNAs (miRNA), represent 
an entirely novel agent of gene regulation 
discovered relatively recently \cite{Ambros, Bartel}, and
have been proposed to function as canalizing agents that buffer the effects of such stochasticity in gene expression
\cite{Stark, Hornstein}. According to this theory, 
when miRNA expression is perturbed, stochasticity in gene expression can result in transitions to 
distinct cellular phenotypes. As miRNAs bind to gene targets they
downregulate translation of target mRNA into protein. Embedded in several different
types of so-called feedforward loops (FFLs),
miRNAs help smooth out noise and generate canalizing effects in gene regulation by overriding
the effect of certain genes on others \cite{MacNeil}.  Complex networks
(viewed as graphs) ranging from the transcriptional networks in yeast and \textit{E. coli} to engineered systems are enriched
for certain graph-theoretic motifs that include feedforward loops \cite{AlonPNAS}. 

An understanding of canalization in evolutionary biology is important as a 
cornerstone of
natural selection and the emergence of new phenotypes \cite{Dassow}, as well as for the
understanding of diseases such as cancer. Transitions to new phenotypes
have been implicated as one of the driving forces of tumorigenesis \cite{huang2012tumor, Kaneko, Capp, Laforge, kauffman1971differentiation}; 
and, interestingly,
significantly altered expression of miRNAs is a feature of several cancers.
Much experimental work remains to be done in elucidating this concept, and recent years have seen
an explosive growth of publications in this area. There have also been a number of computational studies
focused on canalization. Several papers have studied computational models that capture the 
evolution of canalization in networks and their ability to support significant mutation without change in the
phenotype \cite{Bassler, Durrett}, while others have studied models of stochastic gene expression
as the internal source of noise in regulatory networks \cite{Kulkarni}. A detailed stochastic model of the
ability of miRNAs to buffer gene expression noise in a feedforward loop has been proposed \cite{Osella},
providing evidence that this type of network motif can in fact play a canalizing role. There is evidence that 
miRNAs do not regulate their target genes directly; rather they act as post-transcriptional regulators by 
reducing the amount of mRNA and by repressing mRNA translation \cite{miRNA_reg}, e.g., 
by binding to the 3'-UTR of a mRNA, which prevents this mRNA from being translated.

In this paper, we carry out a computational study of the ability of the feedforward loop motif to buffer a gene regulatory network 
against intrinsic noise. This is done using stochastic Boolean network models as a computational instantiation of
gene regulatory networks. We introduce a measure on networks that
captures their ``distance-to-deterministic" characteristics in terms of the stability of their attractors. For a given network, we
successively introduce feedforward loops and track the resulting change in dynamics. The results show clearly that the 
feedforward loop motif buffers the network phenotype, in terms of stability of attractors, against perturbations from 
intrinsic noise.

\section{Modeling framework}
\label{sec:sec2}
\subsection{Stochastic Discrete Dynamical Systems}
In this study, the recently developed framework of stochastic discrete dynamical systems (SDDS) \cite{SDDS} 
is used to model gene regulatory networks. This framework is an appropriate set-up to model
the effect of intrinsic noise on network dynamics. 
A stochastic discrete dynamical system in the variables $x_1,\ldots,x_n$, which in this paper represent genes, 
is defined as a collection of $n$ triplets
\begin{equation}F=(f_i,\pa,\pd)_{i=1}^n,\end{equation}
where
\begin{itemize}
\item $f_i: \{0,1\}^n \rightarrow \{0,1\}$ is the update function for $x_i$ for all $i=1,\ldots,n$,
\item $\pa \in (0,1]$ is the activation propensity,
\item $\pd \in (0,1]$ is the degradation propensity.
\end{itemize}
The stochasticity originates from the propensity parameters $\pa$ and $\pd$, 
which should be interpreted as follows: If there would be an activation of $x_i$
at the next time step, 
i.e., $x_i(t)=0$, and $f_i(x_1(t),\ldots, x_n(t)) = 1$, then $x_i(t+1) = 1$ with probability $\pa$. The degradation probability $\pd$ is defined similarly.

All variables are synchronously updated and one time step can be interpreted as the average time needed for transcription 
and translation of the fastest of the genes considered. 
The propensity parameters for this fastest gene will be set to 1, 
and the propensity parameters of genes with slower transcription and translation take proportionately lower values. Thus, this framework can
be interpreted as introducing a very general stochastic sequential update scheme, which also allows for a variable not to be updated at all
at a given step, a generalization of the usual approach in, e.g., \cite{booleannet}.

\subsection{Quantifying Stochasticity}
In a deterministic discrete dynamical system, each initial configuration lies in exactly one basin of attraction. This changes when stochasticity is introduced. Now, from one initial configuration, different attractors may be reached. To measure the degree of stochasticity in particular dynamics, we look at every initial configuration and regard its transition probabilities to the different attractors. 
If each initial configuration only transitions to one attractor, the dynamics are deterministic, whereas lower maximal transition probabilities to attractors lead to proportionately more stochastic dynamics. In our context, the different attractors may be interpreted as different
cellular phenotypes, which makes the connection to phenotypic stability discussed in the introduction \cite{kauffman1971differentiation}. Thus, this computational project
focuses on the stability of attractors under intrinsic noise, as it is affected by the introduction of feedforward loops. 

Based on this idea, we can define the degree of stochasticity in the 
dynamics of an SDDS $F=(f_i,\pa,\pd)_{i=1}^n$. Let $\mathbb{A}(F)$ be the set 
of all attractors of $F$. Then we can calculate the average maximal transition 
probability to an attractor, where all $2^n$ state space configurations are considered and weighted equally, as follows:
\begin{align}
\mu(F) &= \frac1{2^n} \sum_{x \in \{0,1\}^n}\Big( \underset{A \in \mathbb{A}(F)}{\max } \p (x \overset{F}{\longrightarrow} \cdots \overset{F}{\longrightarrow} A) \Big)\in [0,1].
\end{align}

When $F$ is a deterministic system, $\mu(F)\equiv1$ always holds true. In comparison, for a stochastic system with $a$ attractors, values as low as $1/a$ may be obtained; for stochastic systems with a single attractor, $\mu(F) \equiv 1$ because the single attractor is eventually approached from any initial configuration.

Most limit cycles that are attractors in a deterministic system are no longer attractors in an SDDS, because a cycle can be exited with a certain probability. Nevertheless, one particular kind of limit cycle, which consists of $2^k$ elements and in which all but $k$ bits are fixed, remains an attractor even in a SDDS. One such example is a 2-cycle, in which just one bit switches states, e.g., $000 \leftrightarrow 001$. Table \ref{tab:limit_cycles} shows that such limit cycles occur very rarely by chance, and for computational reasons, we decided to include only steady states and limit cycles of length $2^5$ or less in this study. This restriction does not influence the study since longer limit cycles that remain attractors in the SDDS practically do not occur.

\begin{table}
    \begin{center}
    \caption{This table shows the average number of steady states and limit cycles that remain attractors in the SDDS framework for different network sizes. We found fewer than thirty such 8-cycles among more than 250,000 networks of different sizes, and no 16-cycles at all. This shows that including only limit cycles of length 16 and less is not restricting the study.}
        \begin{tabular}{l|cccc} \hline
Network Size&\multirow{2}{*}{\qquad 5\qquad}&\multirow{2}{*}{\qquad15\qquad}&\multirow{2}{*}{\qquad30\qquad }&\multirow{2}{*}{\qquad50\qquad}\\ \cline{1-1}
Cycle Length & & & & \\ \hline
1& 2.8351&3.6577 &  4.3492 & 4.8709  \\ \hline
2& 0.1529  & 0.1522 & 0.1540& 0.1486\\ \hline
4& 0.0244  & 0.0378& 0.0415& 0.0415\\ \hline
8& 0.0002 & 0& $<$0.0001& $<$0.0001\\ \hline
16& 0  & 0& 0& 0\\ \hline
Sample Size & 120000 & 42000 & 40000 & 62500 \\ \hline
        \end{tabular}
        \label{tab:limit_cycles}
    \end{center}
\end{table}

The basic procedure underlying the computational study is, for a given SDDS, referred to as the ``basic" network,
to construct several ``extended" networks by successively adding nodes, which we shall refer to as miRNAs, together with one or more feedforward loops involving the new
miRNAs in a specific way. We then measure the change in the stochasticity measure described above.
Let $F=(f_i,\pa,\pd)_{i=1}^{n_1}$ be the basic network, and let $F^*=(f^*_i,q_i^{\uparrow},q_i^{\downarrow})_{i=1}^{n_2}$ be the extended network, $n_1\leq n_2$. 
We hypothesize that the dynamics in the feedforward loop enriched network are less stochastic. 
To compare the dynamics of two systems with respect to their degree of determinism, we consider their difference $m$ in $\mu$-values
\begin{equation}
m(F,F^*) = \begin{cases}0 & \text{if } \mu(F^*), \mu(F) = 1\\
\frac{\mu(F^*) - \mu(F)}{1-\min(\mu(F),\mu(F^*))} & \text{otherwise}.\end{cases}
\end{equation}
The denominator scales this difference into the range $[-1,1]$. Here, 
$m=0$ means that both networks have equally stochastic dynamics. 
If $m$ is positive, the extended network $F^*$ is dynamically less stochastic 
than the basic network $F$, and a negative value of $m$ suggests the opposite. 
The magnitude describes the difference in degree of stochasticity. A magnitude of 
$1$ means that one of the networks has the dynamics of a deterministic system, 
whereas a magnitude of $0.5$ suggests that one system is $50\%$ less stochastic than the other. 


\section{Methods}
For each set of input nodes we generated a certain number $N$ of basic Boolean SDDS $F$, introduced miRNAs, in a way that will be specified later in this section, to obtain the extended network $F^*$, and then compared their degree of determinism via $m(F,F^*)$.
We considered 4 network sizes $n$: $5, 15, 30, 50$ nodes. The corresponding number $N$ is $20000, 7000, 5500, 5500$, respectively. 
The networks generated are random, subject to the following restrictions.

Large-scale studies of \textit{B. subtilis}, \textit{E. coli} and \textit{S. cerevisiae} strongly suggest that the in-degrees of nodes in a transcriptional regulatory network are Poisson distributed with a mean of about two\cite{RobEvo}. Thus, we chose a Poisson distribution with parameter $\gamma=2$ for the basic network. The only other
restriction is that each gene is regulated by at least one other gene, which raises the average in-degree to approximately $2.2$. 
The regulators of each gene are randomly chosen from the set of all $n$ genes in the network, allowing self-regulation.

All propensity parameters for transcription factors and miRNAs are also randomly chosen from $[0.2,1]$. For computational reasons, the full interval $[0,1]$ is not used since a propensity parameter close to $0$ might strongly decelerate convergence to attractors, 
slowing the performance of the simulation. However, $0.2$ as the lower limit for the propensity parameters 
still allows one process to happen up to five times more frequently than another.
Each gene is regulated by a certain number of genes, depending on its in-degree, and random Boolean functions 
that actually depend on all input variables are used as update functions. 

\begin{table}
    \begin{center}
    \caption{Example of how an additional miRNA is embedded as an input variable in a target gene's update function. The light grey rectangle contains the original random update function, in which $a_1, \ldots, a_4 \in \{0,1\}$ can be any Boolean values, with the sole restriction that the update function must depend on both inputs. Unless both miRNA and transcription factor are present, the miRNA has no influence since there is no TF mRNA that can be degraded or there is no miRNA that can catalyze this degradation. Only if both are present (gray rows), can the miRNA reduce the TF mRNA level to an extent that the target concentration changes because of the miRNA.}
                    \bgroup
\def\arraystretch{0.75}
        \begin{tabular}{!{\vrule width -2pt}c!{\vrule width -4pt}c!{\vrule width -4pt}c!{\vrule width -4pt}!{\vrule width 1pt}!{\vrule width -3.8pt}c!{\vrule width -2pt}} \hhline{----}\noalign{\vskip+0.2pt}
\ \ miRNA(t) \ \ & \ \ \  \cellcolor[gray]{0.98}TF(t)\ \ \ & \cellcolor[gray]{0.98}{other input(t)}\ \ &\ \ \cellcolor[gray]{0.98}target(t+1)\ \ \\ \hhline{----} \noalign{\vskip+0.2pt}
0 & \cellcolor[gray]{0.98}0 & \cellcolor[gray]{0.98}{0} &\ \ \cellcolor[gray]{0.98}$a_1$ \ \ \\ \hhline{----} \noalign{\vskip+0.2pt}
0 & \cellcolor[gray]{0.98}0 & \cellcolor[gray]{0.98}{1} &\ \ \cellcolor[gray]{0.98}$a_2$ \ \ \\ \hhline{----} \noalign{\vskip+0.2pt}
0 & \cellcolor[gray]{0.98}1 & \cellcolor[gray]{0.98}{0} &\ \ \cellcolor[gray]{0.98}$a_3$ \ \ \\ \hhline{----} \noalign{\vskip+0.2pt}
0 &\cellcolor[gray]{0.98}1 & \cellcolor[gray]{0.98}{1} &\ \ \cellcolor[gray]{0.98}$a_4$ \ \ \\ \hhline{----} \noalign{\vskip+0.2pt}
1 &0 & 0 &\ \  $a_1$ \ \ \\ \hhline{----} \noalign{\vskip+0.2pt}
1 &0 & 1 &\ \  $a_2$ \ \ \\ \hhline{----}\noalign{\vskip+0.2pt}
\rowcolor[gray]{0.8}{1}&{1}& 0 &\ \ {$a_1$} \ \ \\ \hhline{----}\noalign{\vskip+0.2pt}
\rowcolor[gray]{0.8}{1}&{1}& 1 &\ \ {$a_2$} \ \ \\ \hhline{----}\noalign{\vskip+0.2pt}
        \end{tabular}
        \egroup      
        \label{tab:truth_table}
    \end{center}
\end{table}
    
After creating the basic network, we generate an extended network in a way reminiscent of post-transcriptional regulation by miRNAs. (Since we do not include a corresponding protein node for each gene node, the analogy is limited). Starting with the basic SDDS, miRNAs are iteratively added to this system by randomly choosing one gene as a transcription factor (TF) that induces transcription of the introduced miRNA. This miRNA, in turn, reduces the mRNA level of its own transcription factor; one example for such coregulation is the interplay between miR-133b and Pitx3 in midbrain dopamine neurons\cite{miRNA_TF_feedback}. A lower TF mRNA level leads to a lower TF protein level, which then affects the regulation of all TF target genes. In the Boolean framework, a gene-specific threshold is used to distinguish between low (0) and high concentration (1). For some target genes, even after the TF mRNA reduction, there might still be enough transcription factor so that the target concentration is on the same side of the threshold as if no reduction had taken place; for other target genes, the target concentration might change significantly because of the TF mRNA reduction. Since this reduction is caused by the miRNA presence, the miRNA becomes a new regulator of the target genes in the latter case. One input variable in this study, called miRNA strength, describes the probability that transcription factor-target gene pairs fall into this latter case, i.e., that the TF mRNA level is significantly reduced, so that the Booleanized target concentration is the same as if no transcription factor had been present at all. If, for instance, the miRNA strength is 0.5, any miRNA regulates on average half of its transcription factor's target genes. However, we require each miRNA to regulate at least one target gene. This restriction ensures that each miRNA is part of at least one feedforward loop, consisting of transcription factor, miRNA, and target gene. Table \ref{tab:truth_table} depicts an example of how the update function of target genes regulated by a miRNA is expanded, taking into consideration the mode of action of miRNAs. Only if miRNA is present and TF mRNA is transcribed, the mRNA reduction takes place. In this case we see the same output as if no TF mRNA were present.

Because all networks with exactly one attractor already have totally deterministic dynamics, in the sense defined earlier, i.e., $\mu \equiv 1$ for those networks, only basic networks with multiple attractors are considered in this study. Those are the interesting networks, in which actual stabilization of the dynamics might be observed. Particularly interesting dynamics occur if at least two attractors possess a relatively large basin of attraction. The network selection process therefore favors networks with multiple large basins of attraction by picking a network only if at least two attractors are found more than once starting from twenty random initial configurations.

If the extended network $F^*$ does not possess multiple attractors, $m(F,F^*) =1$ by definition. One could argue that the loss of attractors in the extended network is one feature of stabilization through feedforward loops. On the other hand, however, this could be seen as an experimental bias. To consider both views, we use $m$ to define two output measures, $m_1$ and $m_2$, one regarding any extended network and the other considering only those network pairs in which the extended network also possesses multiple attractors with at least two large basins of attraction. For a given set of input variables, we generate $N$ basic and extended networks, and measure
\begin{align}
m_1 &= \frac 1N \sum_{i=1}^N \{m(F,F^*) \big| \text{basic network $F$ and extended network $F^*$ have multiple attractors}\},\\
m_2 &=  \frac 1N \sum_{i=1}^N\{m(F,F^*) \big| \text{only the basic network $F$ is required to have multiple attractors}\}.
\end{align}
For any set of input variables, we expect $m_2 \geq m_1$ since all network pairs with less than two attractors in the extended network, which are omitted in $m_1$, have $\mu \equiv 1$ and thus a mean value closer to 1. However, we do not want to prefer one or the other measure and thus we report results for both, which have been obtained independently, i.e., a network pair that was used for $m_1$ is not used for $m_2$.

A full analysis of the state space of a SDDS is only possible for small networks, so we used random sampling of initial configurations and an estimate of transition probabilities to attractors to approximate $m(F,F^*)$. We created a set of $100$ random initial configurations, which were used in both networks to find the transition probabilities to attractors by updating each configuration 50 times, until an attractor was reached. In a small preliminary study, we found that these two values yield a good trade-off between accuracy and efficiency.

\section{Results}
\begin{figure}
  \centering
  \includegraphics[width=\textwidth]{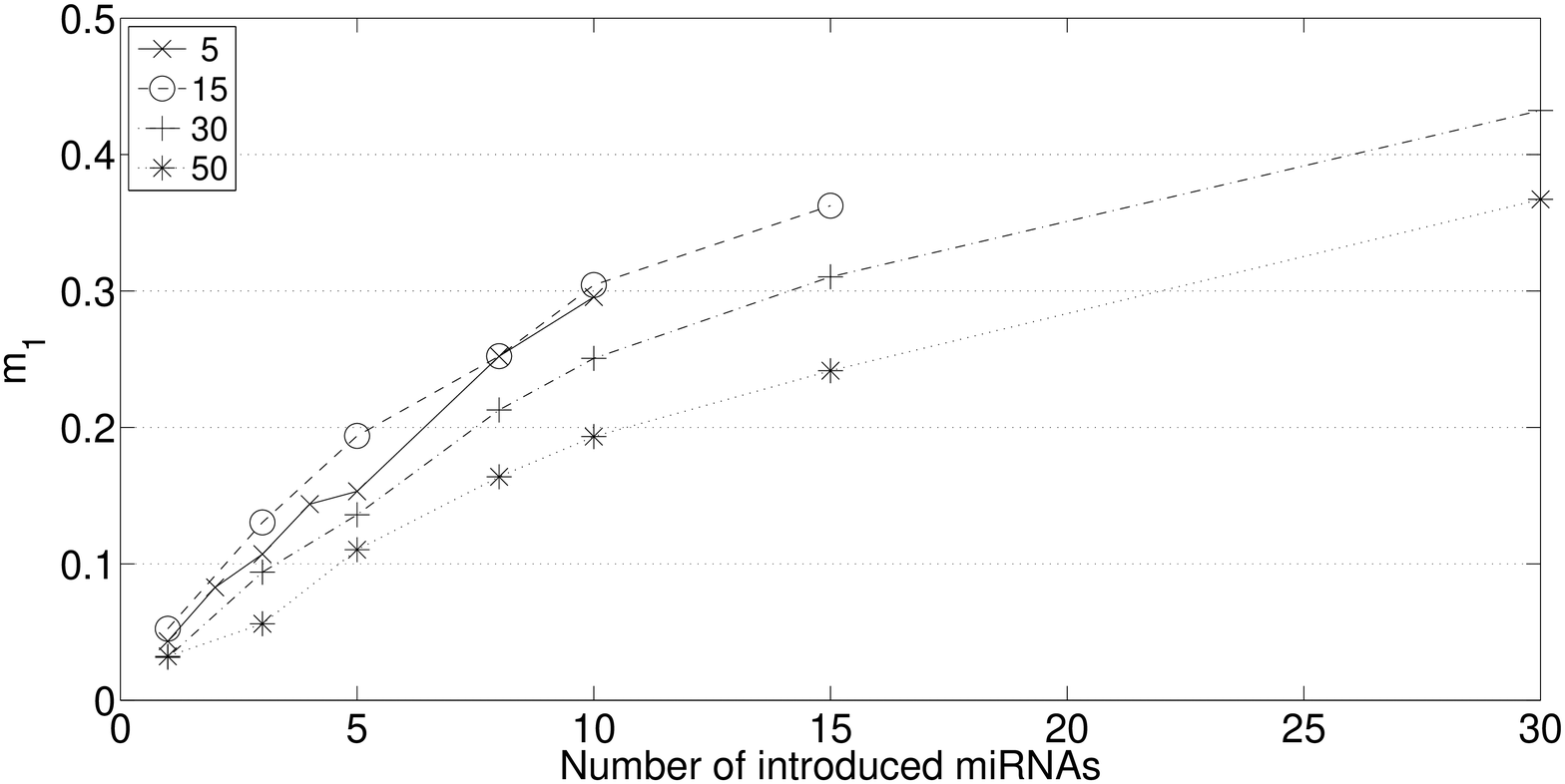}
  \includegraphics[width=\textwidth]{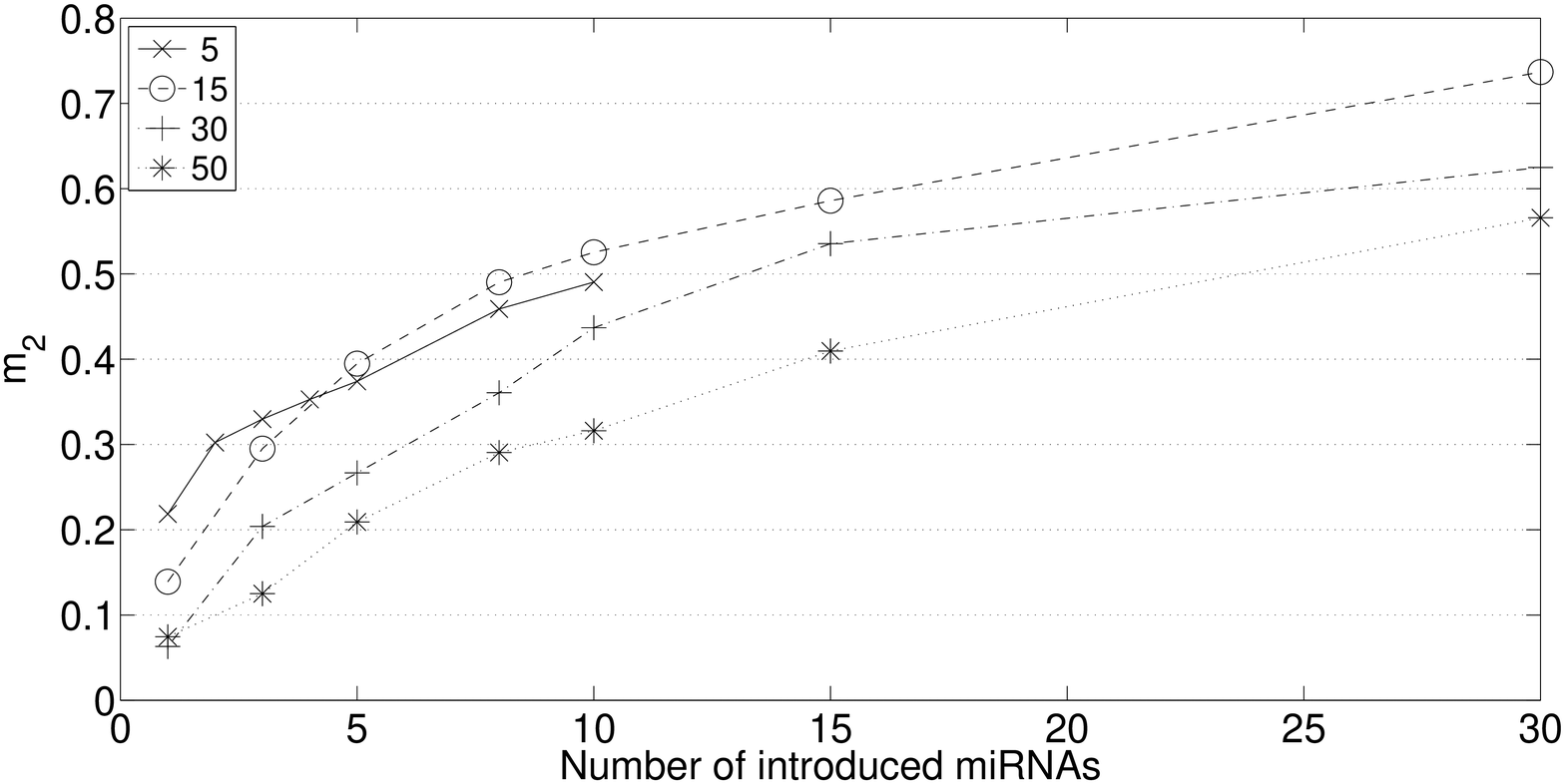}
    \caption{$m_1$ and $m_2$ are plotted against the number of introduced miRNAs. Networks are not necessarily strongly connected, and the miRNA has full strength. The size of the considered networks varies from 5 (solid line) to 50 (dotted line). The impact of a single FFL on the dynamics is larger in smaller networks, which suggests that larger networks require more FFLs for the same amount of stabilization.}
  \label{fig:res1234}
\end{figure}

Overall, we created over $300,000$ pairs of basic and feedforward loop enriched networks. The results for networks with sizes ranging from $5$ to $50$ genes can be seen in Figure \ref{fig:res1234}. None of these networks were required to be strongly connected, and in all of them the introduced miRNAs had full strength, meaning that each miRNA regulates all of its transcription factor's target genes in a feedforward loop structure. The main result is that both measures, $m_1$ and $m_2$, are indeed positive for all network sizes and numbers of introduced miRNAs, indicating that miRNA-mediated FFLs can actually stabilize networks. It can also be seen that the impact of a single miRNA/FFL decreases when the network becomes larger. This means that larger networks require more miRNAs/FFLs for the same degree of stabilization. 

Table \ref{tab:result_n50} shows the results for networks of size 50. We see that more miRNAs and thus more FFLs stabilize the dynamics. Whereas one miRNA of full strength with $m_1 \approx 4\%$ only has a small impact, five such miRNAs already lead to $m_1 \approx 12\%$, and the introduction of thirty miRNAs of full strength stabilizes the stochastic system quite a lot ($m_1 \approx 0.37\%$). As expected, $m_2$ yields higher values and thirty miRNAs already reduce the stochasticity in the dynamics by more than $50\%$ ($m_2 \approx 0.57\%$). In the case that each miRNA regulates on average only half its transcription factor's target genes (but at least one), all values are considerably lower; the general behavior does not change, however.

\begin{table}
    \begin{center}
            \caption{Comparison of the degree of stochasticity via $m_1$ and $m_2$ for not necessarily strongly connected networks of 50 genes, in which various numbers of miRNAs with full strength (Part a) and with strength 0.5 (Part b) are introduced. Overall, the more miRNA-mediated FFLs are introduced, the less stochastic the network dynamics become.}      
        \flushleft{\hskip0.8in \textbf{a)}}
            
            \centering
        \begin{tabular}{l|cccccccc} \hline
Number of miRNAs & 1  & 3 & 5 & 8 & 10 & 15 & 30 \\ \hline
Average Number of FFLs & 2.401  & 7.298  & 12.36  & 19.79 &  24.77 &  37.27 & 74.50\\ \hline
$m_1$ & 0.0374     &   0.0700  &  0.1083  &  0.1548  &  0.1756  &  0.2609 &  0.4042\\ \hline
$m_2$ & 0.0687    &   0.1444 &   0.2178 &   0.2832  &  0.3203  &  0.4297 &   0.5967\\ \hline
        \end{tabular}        
            \vskip3ex
		\flushleft{\hskip0.8in \textbf{b)}}
            
            \centering      
        \begin{tabular}{l|cccccccc} \hline
Number of miRNAs & 1  & 3 & 5 & 8 & 10 & 15 & 30 \\ \hline
Average Number of FFLs & 1.51  &  4.56 &   7.65  & 12.32  & 15.45 &  23.21  & 46.48\\ \hline
$m_1$ & 0.0186   &  0.0523 &  0.0920  &  0.1064  &   0.1356   &  0.1653  &  0.2949\\ \hline
$m_2$ & 0.0441   &  0.1138 &   0.1653 &   0.2052  &  0.2536   & 0.3111    &   0.4682\\ \hline
        \end{tabular}        
        
        \label{tab:result_n50}
    \end{center}
\end{table}

These results raise the question whether a given network can be fully stabilized by introducing a sufficient number of miRNAs. Indeed, under certain conditions this is possible by ensuring the existence of a unique steady state. If an $n$-gene network contains no self-regulating genes, then $n$ miRNAs with full strength, each regulated by another gene, suffice to have fully deterministic network dynamics. Since the miRNA has full strength, it will downregulate any present mRNA, which ensures that only the value $a_1 \in \{0,1\}$ (compare Table \ref{tab:truth_table2}) can be taken by the target gene at a steady state. Each gene is regulated by at least one other gene. Hence, each gene and its regulated miRNA can only take the value $a_1$ in its truth table and the existence of a unique steady state is guaranteed. Thus, $\mu \equiv 1$ for such networks, which is equivalent to fully deterministic dynamics in the sense defined in Section \ref{sec:sec2}. 

\begin{table}
    \begin{center}
            \caption{If each of a target gene's transcription factor regulates a miRNA that degrades the transcription factor mRNA, then only the fixed value $a_1 \in \{0,1\}$ can be taken on at a steady state because each transcription factor and its regulated miRNA have to take on the same value, $0$ or $1$, at a steady state (gray rows). Here, the light gray rectangle contains the original update function, and $a_1, \ldots, a_4$ are any Boolean values such that the update function depends on both inputs.}
            \bgroup
\def\arraystretch{0.85}
        \begin{tabular}{!{\vrule width -2pt}c!{\vrule width -4pt}c!{\vrule width -4pt}c!{\vrule width -4pt}c!{\vrule width -4pt}!{\vrule width 1pt}!{\vrule width -3.8pt}c!{\vrule width -2pt}} \hline \noalign{\vskip+0.2pt}
\ \ miRNA1(t) \ \ & \ \ miRNA2(t) \ \ & \ \ \ \ \ \cellcolor[gray]{0.98}TF1(t)\ \ \ \ \ & \ \ \ \ \  \cellcolor[gray]{0.98}$TF2(t)$\ \ \ \ \ &\ \ \cellcolor[gray]{0.98}target(t+1)\ \ \\ \hline \noalign{\vskip+0.2pt}
\rowcolor[gray]{0.85}{0} & {0} & {0} & {0} &\ \ {$a_1$} \ \ \\ \hline \noalign{\vskip+0.2pt}
0 & 0 &\cellcolor[gray]{0.98}0 & \cellcolor[gray]{0.98}1 &\ \ \cellcolor[gray]{0.98}$a_2$ \ \ \\ \hline \noalign{\vskip+0.2pt}
0 & 0 &\cellcolor[gray]{0.98}1 & \cellcolor[gray]{0.98} 0 &\ \ \cellcolor[gray]{0.98}$a_3$ \ \  \\ \hline \noalign{\vskip+0.2pt}
0 & 0 &\cellcolor[gray]{0.98}1 & \cellcolor[gray]{0.98}1 &\ \ \cellcolor[gray]{0.98}$a_4$ \ \ \\ \hline \noalign{\vskip+0.2pt}
0 & 1 &0 & 0 &\ \  $a_1$ \ \ \\ \hline \noalign{\vskip+0.2pt}
\rowcolor[gray]{0.85}{0} & {1} & {0} & {1} &\ \  {$a_1$} \ \ \\ \hline
0 & 1 &1 & 0 &\ \ $a_3$ \ \ \\ \hline
0 & 1 & 1 & 1 &\ \  $a_3$ \ \ \\ \hline
1 &0 &0 & 0 &\ \ $a_1$ \ \  \\ \hline
1 &0 &0 & 1 &\ \ $a_2$ \ \ \\ \hline \noalign{\vskip+0.2pt}
\rowcolor[gray]{0.85}{1} & {0} & {1} & {0} &\ \  {$a_1$} \ \ \\ \hline
1 &0 &1 & 1 &\ \  $a_2$ \ \  \\ \hline
1 &1 &0 & 0 &\ \  $a_1$ \ \ \\ \hline
1 &1 &0 & 1 &\ \  $a_1$ \ \ \\ \hline
1 &1 &1 & 0 &\ \ $a_1$ \ \ \\ \hline \noalign{\vskip+0.2pt}
\rowcolor[gray]{0.85}{1} & {1} & {1} & {1} &\ \  {$a_1$} \ \ \\ \hline 
        \end{tabular}
        \egroup
        \label{tab:truth_table2}
    \end{center}
\end{table}

\subsection{Derrida Values}
In this study, we introduce a new measure for the robustness of stochastic networks by quantifying the degree of determinism of network dynamics. Another measure that can be used in the Boolean context was suggested by Derrida in 1986 \cite{Derrida}. Pairs of initial configurations of fixed Hamming distance are sampled from the entire state space, and their mean normalized Hamming distance, after being updated using update functions and propensity parameters, is defined as the Derrida value for a given initial Hamming distance. Lower Derrida values reflect more stable dynamics. To take time dependencies into account, we also considered the mean Hamming distance after two and three time steps as has been done earlier \cite{Kauff2}. Table \ref{tab:derrida} displays the percent change in Derrida values starting with a basic $50$-gene network and introducing $30$ miRNAs. In all cases, the change is negative, i.e., the Derrida values decreased, indicating that the extended network exhibits more stable dynamics than the basic network, which we observed for different network sizes as well. Thus, another commonly used robustness measure also agrees with our hypothesis, which suggests that our findings are independent of the choice of robustness measure.

\begin{table} \centering
        \caption{Derrida values for initial small disturbances of a Hamming distance up to $5$ were simulated for a basic $50$-gene network and an extended network with an additional $30$ miRNAs of full strength. Multiple time steps were taken into account to consider time dependencies. The table shows that the percent change of the Derrida values from the basic to the extended network is always negative, indicating that our findings do not depend on the choice of the robustness measure.}
\begin{tabular}{c|ccccc} \hline
Hamming Distance&\multirow{2}{*}{\qquad 1\qquad }&\multirow{2}{*}{\qquad 2\qquad }&\multirow{2}{*}{\qquad 3\qquad }&\multirow{2}{*}{\qquad 4\qquad }&\multirow{2}{*}{\qquad 5\qquad }\\ \cline{1-1}
Time Steps & & & & & \\ \hline
1&   -2.8343  & -5.5589 &  -7.7702 &  -9.6424 & -11.1946 \\ \hline
2&  -3.6709  & -5.0620  & -6.3319  & -7.3764  & -8.3597 \\ \hline
3&   -6.4927 &  -7.4783 &  -8.3548 &  -9.0794  & -9.8100 \\ \hline
        \end{tabular}
        \label{tab:derrida}
        \end{table}

\section{Discussion}
We have examined the effect of feedforward loop motifs in stochastic Boolean network models of transcriptional networks, 
in analogy to the regulatory effects of miRNAs. Our goal was to test the hypothesis that these regulatory motifs have
the effect of buffering the network against stochastic effects in the sense that they stabilize the basins of attraction. To this end, we conducted a computational experiment on a large number of randomly generated networks.
The networks were modified by introducing additional nodes and feedforward motifs in a way that suggests regulation by miRNAs.
To capture the effect on network stability we introduced a new measure of stochasticity of a network suitable for this purpose. 
Using this measure, as well as the classical measure of Derrida values, we showed that indeed the introduction
of miRNAs has the hypothesized buffering effect. 

The number of miRNAs that are introduced strongly influences the magnitude of the stabilizing effect, so that one might wonder how many feedforward loops can be expected to be found in actual gene regulatory networks. In a data set from \textit{E. Coli}, among 424 nodes with 519 edges, 40 FFLs have been found \cite{FFL_EColi}. In \textit{S. cerevisiae}, among 685 genes with 1,052 interactions, there are at least 70 FFLs \cite{FFL_yeast}. However, restricting the data to subnetworks, we find other occurrence frequencies of FFLs. A subnetwork of \textit{E. Coli} of 67 nodes with 102 edges containing 42 FFLs was identified (some new FFLs had been found by then), and in \textit{D. melanogaster}, a subnetwork of 54 nodes and 167 edges contained as many as 157 FFLs \cite{FFL_high}. These numbers indicate that the question of how many FFLs are reasonable in a gene network of a certain size seems to depend strongly on the average in-degree of the nodes; whereas even large networks with average in-degree of less than 2 have few FFLs, this number can rise quickly when the network becomes more highly connected, as indicated by  the considered network of \textit{D. melanogaster}, with an average in-degree of approximately 3.

\begin{figure}
  \centering
  \includegraphics[width=\textwidth]{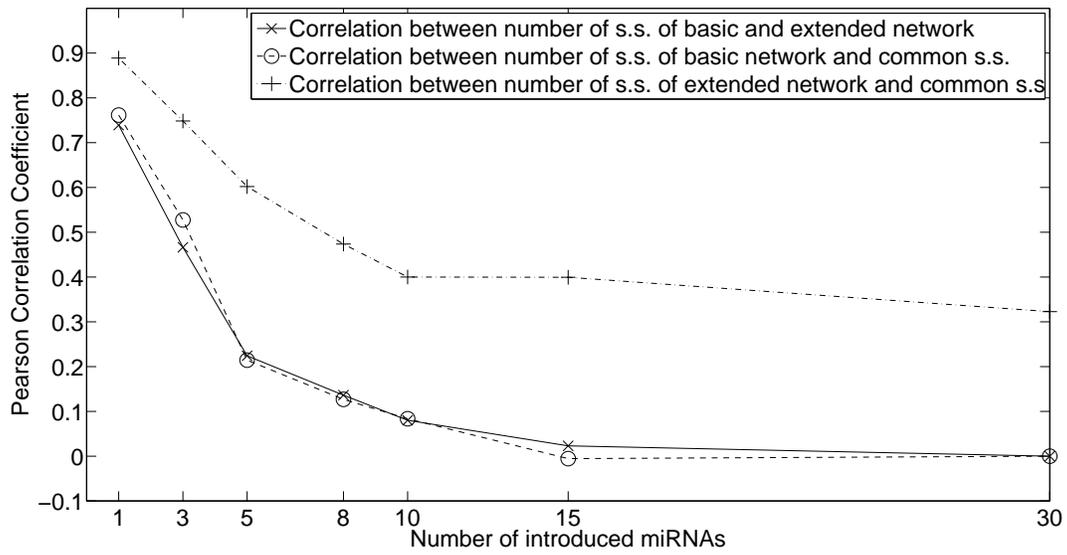}
    \caption{The correlation between the number of attractors in the basic and the extended network and the number of common attractors is compared pairwise. These values have been obtained from not necessarily strongly connected networks of 50 genes and the introduced miRNAs had full strength.}
  \label{fig:cor1}
\end{figure}

Additionally, we looked at the correlations between the number of attractors in both networks and the number of common attractors, where we defined a configuration in both networks to equal if the states of all genes, i.e., the first $n$ bits, coincide. Figure \ref{fig:cor1} shows the observed correlations, and we notice expected decreasing correlations between all three variables when more miRNAs are introduced. Surprisingly, the number of attractors of the extended network is much more strongly correlated with the number of common attractors than the respective number for the basic network, the cause of which remains to be explored.

This study can be extended in several ways, which we are planning to pursue. To make the study design more realistic, it is
useful to introduce additional nodes for proteins, in order to be able to implement more mechanistic details of miRNA regulation. 
Also, here we do not restrict the regulatory rules to those that correspond to activation and inhibition only, which does not allow the
classification of feedforward loops into coherent and incoherent, an important distinction. Also, a more careful study remains
to be done on the effect of miRNAs relative to their position in the network and the local network topology into which they
are embedded. Finally, another limitation of this work is that only intrinsic noise is being considered as a perturbation. It is important,
however, to also take extrinsic noise into account, which requires an extension of the SDDS framework.

\section{Conclusions}
This study provides computational evidence that miRNA-mediated feedforward loops have the effect of buffering
the network against phenotypic variation due to stochastic effects. Introducing a feedforward loop motif has a local effect
on network dynamics that propagates to a generally much smaller global effect on attractor stability. Thus, as the number of
feedforward loop motifs increases, the overall stabilizing effect increases as well. In our study, the number of miRNAs introduced
is of a relative order of magnitude that might be expected in an actual transcriptional network. Thus, our computational experimental
setup can be used in conjunction with an appropriate experimental system to investigate the effects of individual miRNA actions. 

\section{Acknowledgments}
This work was supported by the National Science Foundation under Grant Nr. CMMI-0908201.\\

The computational results presented here were in part obtained using Virginia Tech's Advanced Research Computing (http://www.arc.vt.edu) Ithaca (IBM iDataPlex) system.

\newpage

\begin{thebibliography}{28}%
\makeatletter
\providecommand \@ifxundefined [1]{%
 \@ifx{#1\undefined}
}%
\providecommand \@ifnum [1]{%
 \ifnum #1\expandafter \@firstoftwo
 \else \expandafter \@secondoftwo
 \fi
}%
\providecommand \@ifx [1]{%
 \ifx #1\expandafter \@firstoftwo
 \else \expandafter \@secondoftwo
 \fi
}%
\providecommand \natexlab [1]{#1}%
\providecommand \enquote  [1]{``#1''}%
\providecommand \bibnamefont  [1]{#1}%
\providecommand \bibfnamefont [1]{#1}%
\providecommand \citenamefont [1]{#1}%
\providecommand \href@noop [0]{\@secondoftwo}%
\providecommand \href [0]{\begingroup \@sanitize@url \@href}%
\providecommand \@href[1]{\@@startlink{#1}\@@href}%
\providecommand \@@href[1]{\endgroup#1\@@endlink}%
\providecommand \@sanitize@url [0]{\catcode `\\12\catcode `\$12\catcode
  `\&12\catcode `\#12\catcode `\^12\catcode `\_12\catcode `\%12\relax}%
\providecommand \@@startlink[1]{}%
\providecommand \@@endlink[0]{}%
\providecommand \url  [0]{\begingroup\@sanitize@url \@url }%
\providecommand \@url [1]{\endgroup\@href {#1}{\urlprefix }}%
\providecommand \urlprefix  [0]{URL }%
\providecommand \Eprint [0]{\href }%
\providecommand \doibase [0]{http://dx.doi.org/}%
\providecommand \selectlanguage [0]{\@gobble}%
\providecommand \bibinfo  [0]{\@secondoftwo}%
\providecommand \bibfield  [0]{\@secondoftwo}%
\providecommand \translation [1]{[#1]}%
\providecommand \BibitemOpen [0]{}%
\providecommand \bibitemStop [0]{}%
\providecommand \bibitemNoStop [0]{.\EOS\space}%
\providecommand \EOS [0]{\spacefactor3000\relax}%
\providecommand \BibitemShut  [1]{\csname bibitem#1\endcsname}%
\let\auto@bib@innerbib\@empty
\bibitem [{\citenamefont {Waddington}(1942)}]{wad}%
  \BibitemOpen
  \bibfield  {author} {\bibinfo {author} {\bibfnamefont {C.~H.}\ \bibnamefont
  {Waddington}},\ }\bibfield  {title} {\enquote {\bibinfo {title} {Canalisation
  of development and the inheritance of acquired characters},}\ }\href@noop {}
  {\bibfield  {journal} {\bibinfo  {journal} {Nature}\ }\textbf {\bibinfo
  {volume} {150}},\ \bibinfo {pages} {563--564} (\bibinfo {year}
  {1942})}\BibitemShut {NoStop}%
\bibitem [{\citenamefont {Avigdor}\ and\ \citenamefont
  {Elowitz}(2010)}]{Elowitz}%
  \BibitemOpen
  \bibfield  {author} {\bibinfo {author} {\bibfnamefont {E.}~\bibnamefont
  {Avigdor}}\ and\ \bibinfo {author} {\bibfnamefont {M.}~\bibnamefont
  {Elowitz}},\ }\bibfield  {title} {\enquote {\bibinfo {title} {Functional
  roles for noise in genetic circuits},}\ }\href@noop {} {\bibfield  {journal}
  {\bibinfo  {journal} {Nature}\ }\textbf {\bibinfo {volume} {467}},\ \bibinfo
  {pages} {167--173} (\bibinfo {year} {2010})}\BibitemShut {NoStop}%
\bibitem [{\citenamefont {Ambros}(2004)}]{Ambros}%
  \BibitemOpen
  \bibfield  {author} {\bibinfo {author} {\bibfnamefont {V.}~\bibnamefont
  {Ambros}},\ }\bibfield  {title} {\enquote {\bibinfo {title} {The functions of
  animal micro{R}{N}{A}s},}\ }\href@noop {} {\bibfield  {journal} {\bibinfo
  {journal} {Nature}\ }\textbf {\bibinfo {volume} {431}},\ \bibinfo {pages}
  {350--355} (\bibinfo {year} {2004})}\BibitemShut {NoStop}%
\bibitem [{\citenamefont {Bartel}(2009)}]{Bartel}%
  \BibitemOpen
  \bibfield  {author} {\bibinfo {author} {\bibfnamefont {D.}~\bibnamefont
  {Bartel}},\ }\bibfield  {title} {\enquote {\bibinfo {title} {Micro{R}{N}{A}s:
  target recognition and regulatory functions},}\ }\href@noop {} {\bibfield
  {journal} {\bibinfo  {journal} {Cell}\ }\textbf {\bibinfo {volume} {136}},\
  \bibinfo {pages} {215--233} (\bibinfo {year} {2009})}\BibitemShut {NoStop}%
\bibitem [{\citenamefont {Stark}\ \emph {et~al.}(2005)\citenamefont {Stark},
  \citenamefont {Brennecke}, \citenamefont {Bushati}, \citenamefont {Russell},\
  and\ \citenamefont {Cohen}}]{Stark}%
  \BibitemOpen
  \bibfield  {author} {\bibinfo {author} {\bibfnamefont {A.}~\bibnamefont
  {Stark}}, \bibinfo {author} {\bibfnamefont {J.}~\bibnamefont {Brennecke}},
  \bibinfo {author} {\bibfnamefont {N.}~\bibnamefont {Bushati}}, \bibinfo
  {author} {\bibfnamefont {R.}~\bibnamefont {Russell}}, \ and\ \bibinfo
  {author} {\bibfnamefont {S.}~\bibnamefont {Cohen}},\ }\bibfield  {title}
  {\enquote {\bibinfo {title} {Animal micro{R}{N}{A}s confer robustness to gene
  expression and have a significant impact on 3'{U}{T}{R} evolution},}\
  }\href@noop {} {\bibfield  {journal} {\bibinfo  {journal} {Cell}\ }\textbf
  {\bibinfo {volume} {123}},\ \bibinfo {pages} {1133--1146} (\bibinfo {year}
  {2005})}\BibitemShut {NoStop}%
\bibitem [{\citenamefont {Hornstein}\ and\ \citenamefont
  {Shomron}(2006)}]{Hornstein}%
  \BibitemOpen
  \bibfield  {author} {\bibinfo {author} {\bibfnamefont {E.}~\bibnamefont
  {Hornstein}}\ and\ \bibinfo {author} {\bibfnamefont {N.}~\bibnamefont
  {Shomron}},\ }\bibfield  {title} {\enquote {\bibinfo {title} {Canalization of
  development by micro{R}{N}{A}s},}\ }\href@noop {} {\bibfield  {journal}
  {\bibinfo  {journal} {Nature Genetics}\ }\textbf {\bibinfo {volume} {38}},\
  \bibinfo {pages} {S20--24} (\bibinfo {year} {2006})}\BibitemShut {NoStop}%
\bibitem [{\citenamefont {MacNeil}\ and\ \citenamefont
  {Walhout}(2011)}]{MacNeil}%
  \BibitemOpen
  \bibfield  {author} {\bibinfo {author} {\bibfnamefont {L.}~\bibnamefont
  {MacNeil}}\ and\ \bibinfo {author} {\bibfnamefont {A.}~\bibnamefont
  {Walhout}},\ }\bibfield  {title} {\enquote {\bibinfo {title} {Gene regulatory
  networks and the role of robustness and stochasticity in the control of gene
  expression},}\ }\href@noop {} {\bibfield  {journal} {\bibinfo  {journal}
  {Genome Res}\ }\textbf {\bibinfo {volume} {21}},\ \bibinfo {pages} {645--657}
  (\bibinfo {year} {2011})}\BibitemShut {NoStop}%
\bibitem [{\citenamefont {Mangan}\ and\ \citenamefont {Alon}(2003)}]{AlonPNAS}%
  \BibitemOpen
  \bibfield  {author} {\bibinfo {author} {\bibfnamefont {S.}~\bibnamefont
  {Mangan}}\ and\ \bibinfo {author} {\bibfnamefont {U.}~\bibnamefont {Alon}},\
  }\bibfield  {title} {\enquote {\bibinfo {title} {Structure and function of
  the feed-forward loop network motif},}\ }\href@noop {} {\bibfield  {journal}
  {\bibinfo  {journal} {PNAS}\ }\textbf {\bibinfo {volume} {100}},\ \bibinfo
  {pages} {11980--11985} (\bibinfo {year} {2003})}\BibitemShut {NoStop}%
\bibitem [{\citenamefont {von Dassow}\ and\ \citenamefont
  {Davidson}(2011)}]{Dassow}%
  \BibitemOpen
  \bibfield  {author} {\bibinfo {author} {\bibfnamefont {M.}~\bibnamefont {von
  Dassow}}\ and\ \bibinfo {author} {\bibfnamefont {L.}~\bibnamefont
  {Davidson}},\ }\bibfield  {title} {\enquote {\bibinfo {title} {Physics and
  the canalization of morphogenesis: a grand challenge in organismal
  biology},}\ }\href@noop {} {\bibfield  {journal} {\bibinfo  {journal} {Phys.
  Biol.}\ }\textbf {\bibinfo {volume} {8}} (\bibinfo {year}
  {2011})}\BibitemShut {NoStop}%
\bibitem [{\citenamefont {Huang}(2012)}]{huang2012tumor}%
  \BibitemOpen
  \bibfield  {author} {\bibinfo {author} {\bibfnamefont {S.}~\bibnamefont
  {Huang}},\ }\bibfield  {title} {\enquote {\bibinfo {title} {Tumor
  progression: Chance and necessity in darwinian and lamarckian somatic
  (mutationless) evolution},}\ }\href {\doibase
  10.1016/j.pbiomolbio.2012.05.001} {\bibfield  {journal} {\bibinfo  {journal}
  {Progress in Biophysics and Molecular Biology}\ }\textbf {\bibinfo {volume}
  {110}},\ \bibinfo {pages} {69--86} (\bibinfo {year} {2012})}\BibitemShut
  {NoStop}%
\bibitem [{\citenamefont {Kaneko}(2011)}]{Kaneko}%
  \BibitemOpen
  \bibfield  {author} {\bibinfo {author} {\bibfnamefont {K.}~\bibnamefont
  {Kaneko}},\ }\bibfield  {title} {\enquote {\bibinfo {title} {Characterization
  of stem cells and cancer cells on the basis of gene expression profile
  stability, plasticity, and robustness},}\ }\href@noop {} {\bibfield
  {journal} {\bibinfo  {journal} {Bioessays}\ }\textbf {\bibinfo {volume}
  {33}},\ \bibinfo {pages} {403--413} (\bibinfo {year} {2011})}\BibitemShut
  {NoStop}%
\bibitem [{\citenamefont {Capp}(2011)}]{Capp}%
  \BibitemOpen
  \bibfield  {author} {\bibinfo {author} {\bibfnamefont {J.-P.}\ \bibnamefont
  {Capp}},\ }\bibfield  {title} {\enquote {\bibinfo {title} {Stochastic gene
  expression is the driving force of cancer},}\ }\href@noop {} {\bibfield
  {journal} {\bibinfo  {journal} {Bioessays}\ }\textbf {\bibinfo {volume}
  {33}},\ \bibinfo {pages} {781--782} (\bibinfo {year} {2011})}\BibitemShut
  {NoStop}%
\bibitem [{\citenamefont {Laforge}\ \emph {et~al.}(2005)\citenamefont
  {Laforge}, \citenamefont {Guez}, \citenamefont {Martinez},\ and\
  \citenamefont {Kupiec}}]{Laforge}%
  \BibitemOpen
  \bibfield  {author} {\bibinfo {author} {\bibfnamefont {B.}~\bibnamefont
  {Laforge}}, \bibinfo {author} {\bibfnamefont {D.}~\bibnamefont {Guez}},
  \bibinfo {author} {\bibfnamefont {M.}~\bibnamefont {Martinez}}, \ and\
  \bibinfo {author} {\bibfnamefont {J.}~\bibnamefont {Kupiec}},\ }\bibfield
  {title} {\enquote {\bibinfo {title} {Modeling embryogenesis and cancer: an
  approach based on an equilibrium between autostabilization of stochastic gene
  expression and the interdependence of cells for proliferation},}\ }\href@noop
  {} {\bibfield  {journal} {\bibinfo  {journal} {Prog. Biophys. Mol. Biol.}\
  }\textbf {\bibinfo {volume} {89}},\ \bibinfo {pages} {93--120} (\bibinfo
  {year} {2005})}\BibitemShut {NoStop}%
\bibitem [{\citenamefont {Kauffman}(1971)}]{kauffman1971differentiation}%
  \BibitemOpen
  \bibfield  {author} {\bibinfo {author} {\bibfnamefont {S.}~\bibnamefont
  {Kauffman}},\ }\bibfield  {title} {\enquote {\bibinfo {title}
  {Differentiation of malignant to benign cells},}\ }\href@noop {} {\bibfield
  {journal} {\bibinfo  {journal} {Journal of theoretical biology}\ }\textbf
  {\bibinfo {volume} {31}},\ \bibinfo {pages} {429--451} (\bibinfo {year}
  {1971})}\BibitemShut {NoStop}%
\bibitem [{\citenamefont {Bassler}, \citenamefont {Lee},\ and\ \citenamefont
  {Lee}(2004)}]{Bassler}%
  \BibitemOpen
  \bibfield  {author} {\bibinfo {author} {\bibfnamefont {K.}~\bibnamefont
  {Bassler}}, \bibinfo {author} {\bibfnamefont {C.}~\bibnamefont {Lee}}, \ and\
  \bibinfo {author} {\bibfnamefont {Y.}~\bibnamefont {Lee}},\ }\bibfield
  {title} {\enquote {\bibinfo {title} {Evolution of developmental canalization
  in networks of competing boolean nodes},}\ }\href@noop {} {\bibfield
  {journal} {\bibinfo  {journal} {Phys. Rev. Lett.}\ }\textbf {\bibinfo
  {volume} {93}} (\bibinfo {year} {2004})}\BibitemShut {NoStop}%
\bibitem [{\citenamefont {Huerta-Sanchez}\ and\ \citenamefont
  {Durrett}(2007)}]{Durrett}%
  \BibitemOpen
  \bibfield  {author} {\bibinfo {author} {\bibfnamefont {E.}~\bibnamefont
  {Huerta-Sanchez}}\ and\ \bibinfo {author} {\bibfnamefont {R.}~\bibnamefont
  {Durrett}},\ }\bibfield  {title} {\enquote {\bibinfo {title} {Wagner's
  canalization model},}\ }\href@noop {} {\bibfield  {journal} {\bibinfo
  {journal} {Theor. Popul. Biol.}\ }\textbf {\bibinfo {volume} {71}},\ \bibinfo
  {pages} {121--130} (\bibinfo {year} {2007})}\BibitemShut {NoStop}%
\bibitem [{\citenamefont {JIa}\ and\ \citenamefont
  {Kulkarni}(2010)}]{Kulkarni}%
  \BibitemOpen
  \bibfield  {author} {\bibinfo {author} {\bibfnamefont {T.}~\bibnamefont
  {JIa}}\ and\ \bibinfo {author} {\bibfnamefont {R.}~\bibnamefont {Kulkarni}},\
  }\bibfield  {title} {\enquote {\bibinfo {title} {Post-transcriptional
  regulation of noise in protein distributions during gene expression},}\
  }\href@noop {} {\bibfield  {journal} {\bibinfo  {journal} {Phys. Rev. Lett.}\
  }\textbf {\bibinfo {volume} {105}} (\bibinfo {year} {2010})}\BibitemShut
  {NoStop}%
\bibitem [{\citenamefont {Osella}\ \emph {et~al.}(2011)\citenamefont {Osella},
  \citenamefont {Bosia}, \citenamefont {Cora},\ and\ \citenamefont
  {Caselle}}]{Osella}%
  \BibitemOpen
  \bibfield  {author} {\bibinfo {author} {\bibfnamefont {M.}~\bibnamefont
  {Osella}}, \bibinfo {author} {\bibfnamefont {C.}~\bibnamefont {Bosia}},
  \bibinfo {author} {\bibfnamefont {D.}~\bibnamefont {Cora}}, \ and\ \bibinfo
  {author} {\bibfnamefont {M.}~\bibnamefont {Caselle}},\ }\bibfield  {title}
  {\enquote {\bibinfo {title} {The role of incoherent microrna-mediated
  feedforward loops in noise buffering},}\ }\href@noop {} {\bibfield  {journal}
  {\bibinfo  {journal} {PLoS Comput Biol}\ }\textbf {\bibinfo {volume} {7}}
  (\bibinfo {year} {2011})}\BibitemShut {NoStop}%
\bibitem [{\citenamefont {Valencia-Sanchez}\ \emph {et~al.}(2006)\citenamefont
  {Valencia-Sanchez}, \citenamefont {Liu}, \citenamefont {Hannon},\ and\
  \citenamefont {Parker}}]{miRNA_reg}%
  \BibitemOpen
  \bibfield  {author} {\bibinfo {author} {\bibfnamefont {M.~A.}\ \bibnamefont
  {Valencia-Sanchez}}, \bibinfo {author} {\bibfnamefont {J.}~\bibnamefont
  {Liu}}, \bibinfo {author} {\bibfnamefont {G.~J.}\ \bibnamefont {Hannon}}, \
  and\ \bibinfo {author} {\bibfnamefont {R.}~\bibnamefont {Parker}},\
  }\bibfield  {title} {\enquote {\bibinfo {title} {{{C}ontrol of translation
  and m{R}{N}{A} degradation by mi{R}{N}{A}s and si{R}{N}{A}s}},}\ }\href@noop
  {} {\bibfield  {journal} {\bibinfo  {journal} {Genes Dev.}\ }\textbf
  {\bibinfo {volume} {20}},\ \bibinfo {pages} {515--524} (\bibinfo {year}
  {2006})}\BibitemShut {NoStop}%
\bibitem [{\citenamefont {Murrugarra}\ \emph {et~al.}(2012)\citenamefont
  {Murrugarra}, \citenamefont {Veliz-Cuba}, \citenamefont {Aguilar},
  \citenamefont {Arat},\ and\ \citenamefont {Laubenbacher}}]{SDDS}%
  \BibitemOpen
  \bibfield  {author} {\bibinfo {author} {\bibfnamefont {D.}~\bibnamefont
  {Murrugarra}}, \bibinfo {author} {\bibfnamefont {A.}~\bibnamefont
  {Veliz-Cuba}}, \bibinfo {author} {\bibfnamefont {B.}~\bibnamefont {Aguilar}},
  \bibinfo {author} {\bibfnamefont {S.}~\bibnamefont {Arat}}, \ and\ \bibinfo
  {author} {\bibfnamefont {R.}~\bibnamefont {Laubenbacher}},\ }\bibfield
  {title} {\enquote {\bibinfo {title} {{{M}odeling stochasticity and
  variability in gene regulatory networks}},}\ }\href@noop {} {\bibfield
  {journal} {\bibinfo  {journal} {EURASIP J Bioinform Syst Biol}\ }\textbf
  {\bibinfo {volume} {2012}},\ \bibinfo {pages} {5} (\bibinfo {year}
  {2012})}\BibitemShut {NoStop}%
\bibitem [{\citenamefont {Albert}\ \emph {et~al.}(2008)\citenamefont {Albert},
  \citenamefont {Thakar}, \citenamefont {Li}, \citenamefont {Zhang},\ and\
  \citenamefont {Albert}}]{booleannet}%
  \BibitemOpen
  \bibfield  {author} {\bibinfo {author} {\bibfnamefont {I.}~\bibnamefont
  {Albert}}, \bibinfo {author} {\bibfnamefont {J.}~\bibnamefont {Thakar}},
  \bibinfo {author} {\bibfnamefont {S.}~\bibnamefont {Li}}, \bibinfo {author}
  {\bibfnamefont {R.}~\bibnamefont {Zhang}}, \ and\ \bibinfo {author}
  {\bibfnamefont {R.}~\bibnamefont {Albert}},\ }\bibfield  {title} {\enquote
  {\bibinfo {title} {{{B}oolean network simulations for life scientists}},}\
  }\href@noop {} {\bibfield  {journal} {\bibinfo  {journal} {Source Code Biol
  Med}\ }\textbf {\bibinfo {volume} {3}},\ \bibinfo {pages} {16} (\bibinfo
  {year} {2008})}\BibitemShut {NoStop}%
\bibitem [{\citenamefont {Aldana}\ \emph {et~al.}(2007)\citenamefont {Aldana},
  \citenamefont {Balleza}, \citenamefont {Kauffman},\ and\ \citenamefont
  {Resendiz}}]{RobEvo}%
  \BibitemOpen
  \bibfield  {author} {\bibinfo {author} {\bibfnamefont {M.}~\bibnamefont
  {Aldana}}, \bibinfo {author} {\bibfnamefont {E.}~\bibnamefont {Balleza}},
  \bibinfo {author} {\bibfnamefont {S.}~\bibnamefont {Kauffman}}, \ and\
  \bibinfo {author} {\bibfnamefont {O.}~\bibnamefont {Resendiz}},\ }\bibfield
  {title} {\enquote {\bibinfo {title} {{{R}obustness and evolvability in
  genetic regulatory networks}},}\ }\href@noop {} {\bibfield  {journal}
  {\bibinfo  {journal} {J. Theor. Biol.}\ }\textbf {\bibinfo {volume} {245}},\
  \bibinfo {pages} {433--448} (\bibinfo {year} {2007})}\BibitemShut {NoStop}%
\bibitem [{\citenamefont {Kim}\ \emph {et~al.}(2007)\citenamefont {Kim},
  \citenamefont {Inoue}, \citenamefont {Ishii}, \citenamefont {Vanti},
  \citenamefont {Voronov}, \citenamefont {Murchison}, \citenamefont {Hannon},\
  and\ \citenamefont {Abeliovich}}]{miRNA_TF_feedback}%
  \BibitemOpen
  \bibfield  {author} {\bibinfo {author} {\bibfnamefont {J.}~\bibnamefont
  {Kim}}, \bibinfo {author} {\bibfnamefont {K.}~\bibnamefont {Inoue}}, \bibinfo
  {author} {\bibfnamefont {J.}~\bibnamefont {Ishii}}, \bibinfo {author}
  {\bibfnamefont {W.~B.}\ \bibnamefont {Vanti}}, \bibinfo {author}
  {\bibfnamefont {S.~V.}\ \bibnamefont {Voronov}}, \bibinfo {author}
  {\bibfnamefont {E.}~\bibnamefont {Murchison}}, \bibinfo {author}
  {\bibfnamefont {G.}~\bibnamefont {Hannon}}, \ and\ \bibinfo {author}
  {\bibfnamefont {A.}~\bibnamefont {Abeliovich}},\ }\bibfield  {title}
  {\enquote {\bibinfo {title} {{{A} {M}icro{R}{N}{A} feedback circuit in
  midbrain dopamine neurons}},}\ }\href@noop {} {\bibfield  {journal} {\bibinfo
   {journal} {Science}\ }\textbf {\bibinfo {volume} {317}},\ \bibinfo {pages}
  {1220--1224} (\bibinfo {year} {2007})}\BibitemShut {NoStop}%
\bibitem [{\citenamefont {Derrida}\ and\ \citenamefont
  {Weisbuch}(1986)}]{Derrida}%
  \BibitemOpen
  \bibfield  {author} {\bibinfo {author} {\bibfnamefont {B.}~\bibnamefont
  {Derrida}}\ and\ \bibinfo {author} {\bibfnamefont {G.}~\bibnamefont
  {Weisbuch}},\ }\bibfield  {title} {\enquote {\bibinfo {title} {Evolution of
  overlaps between configurations in random boolean networks},}\ }\href@noop {}
  {\bibfield  {journal} {\bibinfo  {journal} {Journal de physique}\ }\textbf
  {\bibinfo {volume} {47}},\ \bibinfo {pages} {1297--1303} (\bibinfo {year}
  {1986})}\BibitemShut {NoStop}%
\bibitem [{\citenamefont {Kauffman}\ \emph {et~al.}(2003)\citenamefont
  {Kauffman}, \citenamefont {Peterson}, \citenamefont {Samuelsson},\ and\
  \citenamefont {Troein}}]{Kauff2}%
  \BibitemOpen
  \bibfield  {author} {\bibinfo {author} {\bibfnamefont {S.}~\bibnamefont
  {Kauffman}}, \bibinfo {author} {\bibfnamefont {C.}~\bibnamefont {Peterson}},
  \bibinfo {author} {\bibfnamefont {B.}~\bibnamefont {Samuelsson}}, \ and\
  \bibinfo {author} {\bibfnamefont {C.}~\bibnamefont {Troein}},\ }\bibfield
  {title} {\enquote {\bibinfo {title} {Random {B}oolean network models and the
  yeast transcriptional network},}\ }\href@noop {} {\bibfield  {journal}
  {\bibinfo  {journal} {PNAS}\ }\textbf {\bibinfo {volume} {100}},\ \bibinfo
  {pages} {14796--14799} (\bibinfo {year} {2003})}\BibitemShut {NoStop}%
\bibitem [{\citenamefont {Shen-Orr}\ \emph {et~al.}(2002)\citenamefont
  {Shen-Orr}, \citenamefont {Milo}, \citenamefont {Mangan},\ and\ \citenamefont
  {Alon}}]{FFL_EColi}%
  \BibitemOpen
  \bibfield  {author} {\bibinfo {author} {\bibfnamefont {S.~S.}\ \bibnamefont
  {Shen-Orr}}, \bibinfo {author} {\bibfnamefont {R.}~\bibnamefont {Milo}},
  \bibinfo {author} {\bibfnamefont {S.}~\bibnamefont {Mangan}}, \ and\ \bibinfo
  {author} {\bibfnamefont {U.}~\bibnamefont {Alon}},\ }\bibfield  {title}
  {\enquote {\bibinfo {title} {{{N}etwork motifs in the transcriptional
  regulation network of {E}scherichia coli}},}\ }\href@noop {} {\bibfield
  {journal} {\bibinfo  {journal} {Nature Genetics}\ }\textbf {\bibinfo {volume}
  {31}},\ \bibinfo {pages} {64--68} (\bibinfo {year} {2002})}\BibitemShut
  {NoStop}%
\bibitem [{\citenamefont {Milo}\ \emph {et~al.}(2002)\citenamefont {Milo},
  \citenamefont {Shen-Orr}, \citenamefont {Itzkovitz}, \citenamefont {Kashtan},
  \citenamefont {Chklovskii},\ and\ \citenamefont {Alon}}]{FFL_yeast}%
  \BibitemOpen
  \bibfield  {author} {\bibinfo {author} {\bibfnamefont {R.}~\bibnamefont
  {Milo}}, \bibinfo {author} {\bibfnamefont {S.}~\bibnamefont {Shen-Orr}},
  \bibinfo {author} {\bibfnamefont {S.}~\bibnamefont {Itzkovitz}}, \bibinfo
  {author} {\bibfnamefont {N.}~\bibnamefont {Kashtan}}, \bibinfo {author}
  {\bibfnamefont {D.}~\bibnamefont {Chklovskii}}, \ and\ \bibinfo {author}
  {\bibfnamefont {U.}~\bibnamefont {Alon}},\ }\bibfield  {title} {\enquote
  {\bibinfo {title} {{{N}etwork motifs: simple building blocks of complex
  networks}},}\ }\href@noop {} {\bibfield  {journal} {\bibinfo  {journal}
  {Science}\ }\textbf {\bibinfo {volume} {298}},\ \bibinfo {pages} {824--827}
  (\bibinfo {year} {2002})}\BibitemShut {NoStop}%
\bibitem [{\citenamefont {Ishihara}, \citenamefont {Fujimoto},\ and\
  \citenamefont {Shibata}(2005)}]{FFL_high}%
  \BibitemOpen
  \bibfield  {author} {\bibinfo {author} {\bibfnamefont {S.}~\bibnamefont
  {Ishihara}}, \bibinfo {author} {\bibfnamefont {K.}~\bibnamefont {Fujimoto}},
  \ and\ \bibinfo {author} {\bibfnamefont {T.}~\bibnamefont {Shibata}},\
  }\bibfield  {title} {\enquote {\bibinfo {title} {{{C}ross talking of network
  motifs in gene regulation that generates temporal pulses and spatial
  stripes}},}\ }\href@noop {} {\bibfield  {journal} {\bibinfo  {journal} {Genes
  Cells}\ }\textbf {\bibinfo {volume} {10}},\ \bibinfo {pages} {1025--1038}
  (\bibinfo {year} {2005})}\BibitemShut {NoStop}%
\end{thebibliography}
\end{document}